\documentclass[a4paper]{article}
\usepackage[hmargin=2.3cm,vmargin=2.8cm]{geometry}
\usepackage[latin1]{inputenc}
\usepackage[T1]{fontenc}
\usepackage{pgf}
\usepackage[english]{babel}
\usepackage{amsmath,enumerate}
\usepackage{amssymb}
\usepackage{amscd}
\usepackage{tabularx}
\usepackage{tikz}
\usepackage{url}
\usepackage{hyperref}
\usepackage{algorithm,algorithmic}

\newcommand{\set}[1]{\left\{ #1 \right\}}
\newcommand{\abs}[1]{\left| #1 \right|}

\usetikzlibrary{arrows,arrows.meta,shapes,chains,matrix,positioning,scopes}
\usetikzlibrary{calc}
\usetikzlibrary{decorations.pathmorphing}

\newcommand{\EdgeUn}{green!50!black}
\newcommand{\EdgeDeux}{red}
\newcommand{\EdgeTrois}{blue}

\newcommand{\myedge}[4][]{
\draw[thick,#1] #2 edge #3;
\node[circle,inner sep=0.2mm, fill=white] at ($#2!0.5!#3$) {\footnotesize #4};
}

\newcommand{\Un}[2]{\myedge[\EdgeUn]{#1}{#2}{1}}
\newcommand{\Deux}[2]{\myedge[\EdgeDeux]{#1}{#2}{2}}
\newcommand{\Trois}[2]{\myedge[\EdgeTrois]{#1}{#2}{3}}

\usepackage{graphicx} 
\usepackage{subfig}
\usepackage{caption}
\usepackage{soul}

\newenvironment{proof}[1][\unskip]{\par\addvspace{\medskipamount}\noindent\textbf{Proof #1.} \ignorespaces}{\hfill$\Box$ \par\addvspace{\medskipamount}}
\newenvironment{subproof}[1][\unskip]{\par\addvspace{\medskipamount}\noindent\textbf{Proof #1.} \ignorespaces}{\hfill$\blacksquare$ \par\addvspace{\medskipamount}}
\newtheorem{theorem}{Theorem}

\newtheorem{claim}[theorem]{Claim}

\title{Complexity and algorithms for injective edge-coloring in graphs\thanks{This research was supported by the IFCAM project ``Applications of graph homomorphisms'' (MA/IFCAM/18/39) and by the ANR project HOSIGRA (ANR-17-CE40-0022).}}
 \author{Florent Foucaud\footnote{\noindent LIMOS, CNRS UMR 6158, Universit\'e Clermont Auvergne, Aubi\`ere, France.}~\footnote{\noindent Univ. Bordeaux, CNRS, Bordeaux INP, LaBRI, UMR5800, F-33400 Talence, France}~\footnote{\noindent Univ. Orl\'eans, INSA Centre Val de Loire, LIFO EA 4022, F-45067 Orl\'eans, France.}
   \and Herv\'e Hocquard\footnotemark[3]
     \and Dimitri Lajou\footnotemark[3]
   }

\newcommand{\InjPbName}[1]{\textsc{Injective $#1$-Edge-Coloring}}

\begin{document}

\maketitle

\begin{abstract}
An injective $k$-edge-coloring of a graph $G$ is an assignment of colors, i.e. integers in $\set{1, \ldots , k}$, to the edges of $G$ such that any two edges each incident with one distinct endpoint of a third edge, receive distinct colors. The problem of determining whether such a $k$-coloring exists is called \InjPbName{k}. We show that \InjPbName{3} is NP-complete, even for triangle-free cubic graphs, planar subcubic graphs of arbitrarily large girth, and planar bipartite subcubic graphs of girth~6. \InjPbName{4} remains NP-complete for cubic graphs. For any $k\geq 45$, we show that \InjPbName{k} remains NP-complete even for graphs of maximum degree at most $5\sqrt{3k}$. In contrast with these negative results, we show that \InjPbName{k} is linear-time solvable on graphs of bounded treewidth. Moreover, we show that all planar bipartite subcubic graphs of girth at least~16 are injectively $3$-edge-colorable. In addition, any graph of maximum degree at most $\sqrt{k/2}$ is injectively $k$-edge-colorable.
\end{abstract}

\section{Introduction}

We study the algorithmic complexity of the injective edge-coloring problem. Our aim is to determine restricted graph classes where the problem is NP-hard, while in contrast, designing algorithms for other graph classes.
An \emph{injective $k$-edge-coloring} of a graph $G = (V(G), E(G))$ is an assignment of colors, i.e. integers in $\set{1, \ldots , k}$, to the edges of $G$ in such a way that two edges that are each incident with one distinct endpoint of a third edge, receive distinct colors. In other words, for any $3$-edge path of $G$ (possibly forming a triangle), the first and last edge of the path receive distinct colors. The \emph{injective chromatic index} of $G$, denoted $\chi'_i(G)$, is the least integer $k$ for which $G$ admits an injective $k$-edge-coloring.

This concept was recently introduced in~\cite{Cardoso19}, where it is studied for some classes of graphs, and proved to be NP-complete. Bounds on the injective chromatic index of planar graphs, graphs of given maximum degree, and other important graph classes, have been recently determined in~\cite{Axenovich19,Bu18,Raspaud19,Kostochka20,Yue16}. In particular, as mentioned in~\cite{Raspaud19}, it follows from~\cite{Axenovich19} that all planar graphs are injectively $30$-edge-colorable, while outerplanar graphs are injectively $9$-edge-colorable~\cite{Raspaud19}. It is also proved in~\cite{Kostochka20} that subcubic graphs are injectively $7$-edge-colorable, while subcubic bipartite graphs~\cite{Raspaud19} and subcubic planar graphs~\cite{Kostochka20} are injectively $6$-edge-colorable. Moreover all subcubic planar bipartite graphs are injectively $4$-edge-colorable~\cite{Kostochka20}.

Note that in~\cite{Axenovich19}, this notion is studied as the \emph{induced star arboricity} of a graph, that is, the smallest number of star forests into which the edges of the graph can be partitioned: this is an equivalent way to interpret injective edge-coloring (see~\cite{Raspaud19}). The concept of an injective edge-coloring is the natural edge-version of the notion of an injective vertex-coloring, introduced in~\cite{Hahn02} and well-studied since then.% In the case of vertex-colorings, a more general version has been introduced in the book~\cite[Chapter 11.9]{sparsity-book} in the context of the theory of sparse graphs. There, after fixing a positive integer $p$, we wish to color the vertices of the graph in such a way that two vertices connected with a (not necessarily shortest) path of length~$p$, receive distinct colors. For $p=2$, we get the notion of injective vertex-colorings. %Just as proper edge-colorings can be seen as proper vertex-colorings of line graphs, injective edge-colorings can be seen as injective vertex-colorings of line graphs.\todo{Faux!}

Another closely related notion is the one of \emph{strong edge-coloring} of a graph $G$, introduced in~\cite{Fouquet83} and well-studied since then, especially in view of a celebrated conjecture by Erd\H{o}s and Ne\v{s}et\v{r}il~\cite{Erdos88}. In this type of coloring, edges that are the endpoints of a same $3$-edge path \emph{or $2$-edge path} must receive distinct colors. The \emph{strong chromatic index} $\chi_s'(G)$ of a graph $G$ is the least integer $k$ for which $G$ admits a strong edge-coloring with $k$ colors. %, and such colorings can be seen as proper vertex-colorings of squares of line graphs.
It follows from the definitions that for any graph $G$, $\chi'_i(G)\leq \chi'_s(G)$ holds.

The algorithmic complexity of determining the strong chromatic index of a graph is well-studied, see for example~\cite{holyer} for a classic reference, and~\cite{Cole08,Hocquard13} for more recent ones. In this paper, we wish to undertake similar types of studies for the injective chromatic index. The problem at hand is formally defined as follows.

\medskip\noindent
\InjPbName{k} \\
Instance: A graph $G$.\\
Question: Does $G$ admit an injective $k$-edge-coloring?
\medskip

\InjPbName{k} was proved NP-complete (for every fixed $k\geq 3$) in~\cite{Cardoso19}, with no particular restriction on the inputs. We strengthen this as follows.

\begin{theorem}\label{th:cubic}
The two following are NP-Complete:
%1. \InjPbName{3}, even for triangle-free cubic graphs, and\\
%2. \InjPbName{4}, even for cubic graphs.
\begin{enumerate}
    \item \InjPbName{3}, even for triangle-free cubic graphs, and\label{th:3-cubic}
    \item \InjPbName{4}, even for cubic graphs.\label{th:4-cubic}
\end{enumerate}
\end{theorem}

% \begin{theorem}\label{th:cubic}
% \begin{enumerate}
%     \item \InjPbName{3} is NP-Complete even for triangle-free cubic graphs, and\label{th:3-cubic}
%     \item \InjPbName{4} is NP-Complete even for cubic graphs.\label{th:4-cubic}
% \end{enumerate}
% \end{theorem}

%\begin{theorem}\label{th:4-cubic}
%\InjPbName{4} is NP-Complete even for cubic graphs.
%\end{theorem}

Answering a question from~\cite{Cardoso19} about the complexity of \InjPbName{k} for planar graphs, we also study restricted subclasses of planar graphs.

\begin{theorem}\label{th:3-planar}
Let $g \geq 3$. \InjPbName{3} is NP-Complete even for:
\begin{enumerate}
    \item planar subcubic graphs with girth at least $g$,
    \item planar bipartite subcubic graphs of girth 6.
\end{enumerate}
\end{theorem}

The two items in Theorem~\ref{th:3-planar} cannot be combined, because we can prove the following (note that all planar bipartite subcubic graphs are injectively $4$-edge-colorable~\cite{Kostochka20}).

\begin{theorem}\label{th:maille16}
Every planar bipartite subcubic graph of girth at least 16 is injectively $3$-edge-colorable.
\end{theorem}

 We also obtain the following positive result ($tw(G)$ denotes the treewidth of $G$).

\begin{theorem}\label{th:fpt}
For every graph $G$ of order $n$ and every positive integer $k$, there exists a $2^{O(k\cdot tw(G)^2)}n$-time algorithm that solves \InjPbName{k}.
\end{theorem}

It is proved in~\cite{Axenovich19} that $\chi'_i(G)\leq 3{tw(G) \choose 2}$, and so using the above algorithm, one can determine the injective chromatic index of a graph of order $n$ in time $2^{O(tw(G)^4)}n$.

Contrasting with our hardness results for planar graphs, Theorem~\ref{th:fpt} implies that \InjPbName{k} can be solved in polynomial-time on subclasses of planar graphs: $K_4$-minor-free graphs (i.e. graphs of treewidth $2$), and thus, on outerplanar graphs.

%\begin{corollary}\label{cor:outerplanar}
%The problem \InjPbName{p} can be solve in linear time on outer-planar graphs.
%\end{corollary}

%Focusing on the maximum degree, we can prove the following.
%Improving the general reduction 
In \cite{Cardoso19}, Cardoso \textit{et al.} use a reduction on graphs having their maximum degree linear in the number of colors. We improve it with the following result.

\begin{theorem}\label{th:small-Delta-NPC}
For every integer $k \geq 45$, \InjPbName{k} is NP-Complete even for graphs with maximum degree  at most $5\sqrt{3k}$.
\end{theorem}

The bound of Theorem~\ref{th:small-Delta-NPC} is tight up to a constant factor: by a standard maximum degree argument of a conflict graph, every graph with maximum degree at most $\sqrt{k/2}$ is injectively $k$-edge-colorable. (Indeed, for every edge $e$ of a graph $G$, there are at most $2(\Delta(G) -1)^2$ edges which cannot have the same color as $e$, where $\Delta(G)$ is the maximum degree of $G$.)

%\begin{proposition}\label{prop:small-Delta-poly}
%\InjPbName{k} is polynomial-time solvable for graphs with maximum degree at most \sqrt{\tfrac{k}{2}}.
%\end{proposition}

\section{Proof of Theorem~\ref{th:cubic}}
For these two problems, we reduce from \textsc{$3$-Edge-Coloring}, which is NP-Complete even for cubic graphs~\cite{holyer}. (Recall that a proper edge-coloring is an edge-coloring for which edges that are incident to a same vertex receive different colors.)

\medskip\noindent
\textsc{$3$-Edge-Coloring} \\
Instance: A cubic graph $G$.\\
Question: Does $G$ admit a proper $3$-edge-coloring?
\medskip

\subsection{Proof of Theorem~\ref{th:cubic}.\ref{th:3-cubic}}

\begin{proof}
Let $G$ be the input cubic graph. We will proceed in two steps: first, we create a triangle-free subcubic graph $G'$ which has an injective $3$-edge-coloring if and only if $G$ is properly $3$-edge-colorable. Then we describe how to make the graph cubic.

We create the graph $G'$ from $G$ by removing all the edges of $G$. For each edge $uv$ of $G$, we create a copy of a gadget $E_{uv}$ (see Figure~\ref{fig:cubic-3colors-fig}(a) for an illustration) and connect it to $u$ and $v$ as follows. We add eight new vertices $w_{uv}$, $z_{uv}$, $a_{uv}$, $b_{uv}$, $c_{uv}$, $d_{uv}$, $e_{uv}$ and $f_{uv}$. We create the following edges $uw_{uv}$, $vw_{uv}$, $w_{uv}z_{uv}$, $z_{uv}a_{uv}$, $z_{uv}b_{uv}$, $a_{uv}c_{uv}$, $b_{uv}c_{uv}$, $a_{uv}d_{uv}$, $b_{uv}e_{uv}$, $c_{uv}f_{uv}$, $d_{uv}f_{uv}$ and $e_{uv}f_{uv}$.

\begin{claim}\label{cl:3ec-subcubic}
$E_{uv}$ is injectively $3$-edge-colorable, and for every valid edge-coloring $\gamma$ of $E_{uv}$, $\gamma(uw_{uv})=\gamma(vw_{uv})=\gamma(w_{uv}z_{uv})$. Moreover, for any choice of the same color for these three edges, we can extend the coloring to an injective $3$-edge-coloring of $E_{uv}$.
\end{claim}

\begin{subproof}
Let us injectively $3$-edge-color $E_{uv}$. W.l.o.g., we can assume that $d_{uv}f_{uv}$ is colored $1$, $b_{uv}c_{uv}$ is colored $2$ and $a_{uv}z_{uv}$ is colored $3$. We deduce that $b_{uv}e_{uv}$ is colored $2$, $c_{uv}f_{uv}$ is colored $1$, $a_{uv}d_{uv}$ and $a_{uv}c_{uv}$ are colored $3$, $b_{uv}z_{uv}$ is colored $2$ and $e_{uv}f_{uv}$ is colored $1$. Hence $uw_{uv}$, $vw_{uv}$ and $w_{uv}z_{uv}$ must all be colored $1$. 

Now, given one same color for these three edges, one can color the rest of the gadget, for example using the previously constructed coloring. 
\end{subproof}

If $G$ has a proper $3$-edge-coloring $\gamma$, we injectively $3$-edge-color $G'$ by assigning to $uw_{uv}$, $vw_{uv}$ and $w_{uv}z_{uv}$ in $G'$ the color $\gamma(uv)$; then we extend the coloring to each $E_{uv}$ using Claim~\ref{cl:3ec-subcubic}.

Conversely, if $G'$ has an injective $3$-edge-coloring, then we color an edge $uv$ of $G$ with the color of the edge $uw_{uv}$ (or $vw_{uv}$) of $G'$. This coloring is proper since Claim~\ref{cl:3ec-subcubic} insures that $uw_{uv}$ and $vw_{uv}$ have the same color. Indeed if $ux$ is an edge adjacent to $uv$, then $uw_{uv}$ and $xw_{ux}$ have different colors.

\medskip

We now show how to make the construction cubic. We create the cubic graph $G''$ as follows. First, take three disjoint copies $G_1$, $G_2$ and $G_3$ of $G'$. To differentiate the vertices of each copy, we add an exponent to the name of the vertex corresponding to the number of the copy. For example, vertex $w_{uv}$ of $G_1$ will be noted $w_{uv}^1$. For each edge $uv$ of $G$, connect $G_1$, $G_2$ and $G_3$ via $K_{1,3}$  with vertex classes $\set{r_{uv}}$ and $\set{s_{uv}, p_{uv}, q_{uv}}$ as follows. The vertex $s_{uv}$ (resp. $p_{uv}$, resp. $q_{uv}$) is adjacent to $d_{uv}^3$ (resp. $d_{uv}^1$, resp. $d_{uv}^2$), $e_{uv}^2$ (resp. $e_{uv}^3$, resp. $e_{uv}^1$)  and $r_{uv}$ (see Figure~\ref{fig:cubic-3colors-fig}(b)). 
The graph $G''$ is simply the graph where the edge gadget is represented in Figure~\ref{fig:cubic-3colors-fig} and for each $u \in V(G)$, the three copies of $u^i$ for $i \in \set{1,2,3}$ are identified.

As $G$ is cubic, $G''$ is triangle-free and cubic. Note that if $G''$ admits an injective $3$-edge-coloring, then in particular $G'$ also admits an injective $3$-edge-coloring and thus by our previous arguments, $G$ is properly $3$-edge-colorable.

\begin{figure*}[t]
    \centering
    \scalebox{.8}{\begin{tikzpicture}
\tikzstyle{n} = [draw, circle, inner sep=0pt, text width=6mm, align=center];

\node at (0,-5.5) {(a) Edge gadget $E_{uv}$ with an injective $3$-edge-coloring.};
\node at (11,-5.5) {(b) Connecting three copies of $E_{uv}$ in the construction};
\node at (11,-6) {of $G''$, along with an injective $3$-edge-coloring.};

  \begin{scope}[rotate=90,xshift=3cm]
\foreach \i in {1}{
    \node[n] (f\i) at (\i*360/3 +360/6:1.5) {$f_{uv}$}; 
    \node[n] (e\i) at (\i*360/3 +360/6+360/12:2.5) {$e_{uv}$}; 
    \node[n] (d\i) at (\i*360/3 +360/6-360/12:2.5) {$d_{uv}$}; 
    \node[n] (c\i) at (\i*360/3 +360/6:3) {$c_{uv}$}; 
    \node[n] (b\i) at (\i*360/3 +360/6+360/18:4) {$b_{uv}$}; 
    \node[n] (a\i) at (\i*360/3 +360/6-360/18:4) {$a_{uv}$}; 
    \node[n] (z\i) at (\i*360/3 +360/6:4.5) {$z_{uv}$}; 
    \node[n] (w\i) at (\i*360/3 +360/6:6) {$w_{uv}$}; 
    \node[n] (u\i) at (\i*360/3 +360/6+360/18:6) {$u$}; 
    \node[n] (v\i) at (\i*360/3 +360/6-360/18:6) {$v$}; 
    }

\myedge[\EdgeUn]{(u1)}{(w1)}{1}
\myedge[\EdgeUn]{(v1)}{(w1)}{1}
\myedge[\EdgeUn]{(z1)}{(w1)}{1}

\myedge[\EdgeUn]{(f1)}{(d1)}{1}
\myedge[\EdgeUn]{(f1)}{(e1)}{1}
\myedge[\EdgeUn]{(f1)}{(c1)}{1}

\myedge[\EdgeDeux]{(b1)}{(z1)}{2}
\myedge[\EdgeDeux]{(b1)}{(c1)}{2}
\myedge[\EdgeDeux]{(b1)}{(e1)}{2}

\myedge[\EdgeTrois]{(a1)}{(z1)}{3}
\myedge[\EdgeTrois]{(a1)}{(c1)}{3}
\myedge[\EdgeTrois]{(a1)}{(d1)}{3}

\end{scope}

  \begin{scope}[xshift=11cm,rotate=30]
\node[n] (r) at (0,0) {$r_{uv}$};

\node[n] (s) at (0:1.5) {$s_{uv}$};
\node[n] (p) at (360/3:1.5) {$p_{uv}$};
\node[n] (q) at (2*360/3:1.5) {$q_{uv}$};

\foreach \i in {1,2,3}{
    \node[n] (f\i) at (\i*360/3 +360/6:1.5) {$f_{uv}^{\i}$}; 
    \node[n] (e\i) at (\i*360/3 +360/6+360/12:2.5) {$e_{uv}^{\i}$}; 
    \node[n] (d\i) at (\i*360/3 +360/6-360/12:2.5) {$d_{uv}^{\i}$}; 
    \node[n] (c\i) at (\i*360/3 +360/6:3) {$c_{uv}^{\i}$}; 
    \node[n] (b\i) at (\i*360/3 +360/6+360/18:4) {$b_{uv}^{\i}$}; 
    \node[n] (a\i) at (\i*360/3 +360/6-360/18:4) {$a_{uv}^{\i}$}; 
    \node[n] (z\i) at (\i*360/3 +360/6:4.5) {$z_{uv}^{\i}$}; 
    \node[n] (w\i) at (\i*360/3 +360/6:6) {$w_{uv}^{\i}$}; 
    \node[n] (u\i) at (\i*360/3 +360/6+360/18:6) {$u^{\i}$}; 
    \node[n] (v\i) at (\i*360/3 +360/6-360/18:6) {$v^{\i}$}; 
}

%%% G1
\myedge[\EdgeUn]{(u1)}{(w1)}{1}
\myedge[\EdgeUn]{(v1)}{(w1)}{1}
\myedge[\EdgeUn]{(z1)}{(w1)}{1}

\myedge[\EdgeUn]{(f1)}{(d1)}{1}
\myedge[\EdgeUn]{(f1)}{(e1)}{1}
\myedge[\EdgeUn]{(f1)}{(c1)}{1}

\myedge[\EdgeDeux]{(b1)}{(z1)}{2}
\myedge[\EdgeDeux]{(b1)}{(c1)}{2}
\myedge[\EdgeDeux]{(b1)}{(e1)}{2}

\myedge[\EdgeTrois]{(a1)}{(z1)}{3}
\myedge[\EdgeTrois]{(a1)}{(c1)}{3}
\myedge[\EdgeTrois]{(a1)}{(d1)}{3}

%%% G2

\myedge[\EdgeDeux]{(u2)}{(w2)}{2}
\myedge[\EdgeDeux]{(v2)}{(w2)}{2}
\myedge[\EdgeDeux]{(z2)}{(w2)}{2}

\myedge[\EdgeDeux]{(f2)}{(d2)}{2}
\myedge[\EdgeDeux]{(f2)}{(e2)}{2}
\myedge[\EdgeDeux]{(f2)}{(c2)}{2}

\myedge[\EdgeUn]{(a2)}{(z2)}{1}
\myedge[\EdgeUn]{(a2)}{(c2)}{1}
\myedge[\EdgeUn]{(a2)}{(d2)}{1}

\myedge[\EdgeTrois]{(b2)}{(z2)}{3}
\myedge[\EdgeTrois]{(b2)}{(c2)}{3}
\myedge[\EdgeTrois]{(b2)}{(e2)}{3}

%%% G3

\myedge[\EdgeTrois]{(u3)}{(w3)}{3}
\myedge[\EdgeTrois]{(v3)}{(w3)}{3}
\myedge[\EdgeTrois]{(z3)}{(w3)}{3}

\myedge[\EdgeTrois]{(f3)}{(d3)}{3}
\myedge[\EdgeTrois]{(f3)}{(e3)}{3}
\myedge[\EdgeTrois]{(f3)}{(c3)}{3}

\myedge[\EdgeDeux]{(a3)}{(z3)}{2}
\myedge[\EdgeDeux]{(a3)}{(c3)}{2}
\myedge[\EdgeDeux]{(a3)}{(d3)}{2}

\myedge[\EdgeUn]{(b3)}{(z3)}{1}
\myedge[\EdgeUn]{(b3)}{(c3)}{1}
\myedge[\EdgeUn]{(b3)}{(e3)}{1}

%%% Center
\myedge[\EdgeTrois]{(q)}{(d2)}{3}
\myedge[\EdgeTrois]{(q)}{(e1)}{3}
\myedge[\EdgeTrois]{(q)}{(r)}{3}

\myedge[\EdgeDeux]{(p)}{(d1)}{2}
\myedge[\EdgeDeux]{(p)}{(e3)}{2}
\myedge[\EdgeDeux]{(p)}{(r)}{2}

\myedge[\EdgeUn]{(s)}{(d3)}{1}
\myedge[\EdgeUn]{(s)}{(e2)}{1}
\myedge[\EdgeUn]{(s)}{(r)}{1}
\end{scope}
\end{tikzpicture}}
    \caption{Edge gadgets used in the proof of Theorem~\ref{th:cubic}.\ref{th:3-cubic}.}
    \label{fig:cubic-3colors-fig}
\end{figure*}
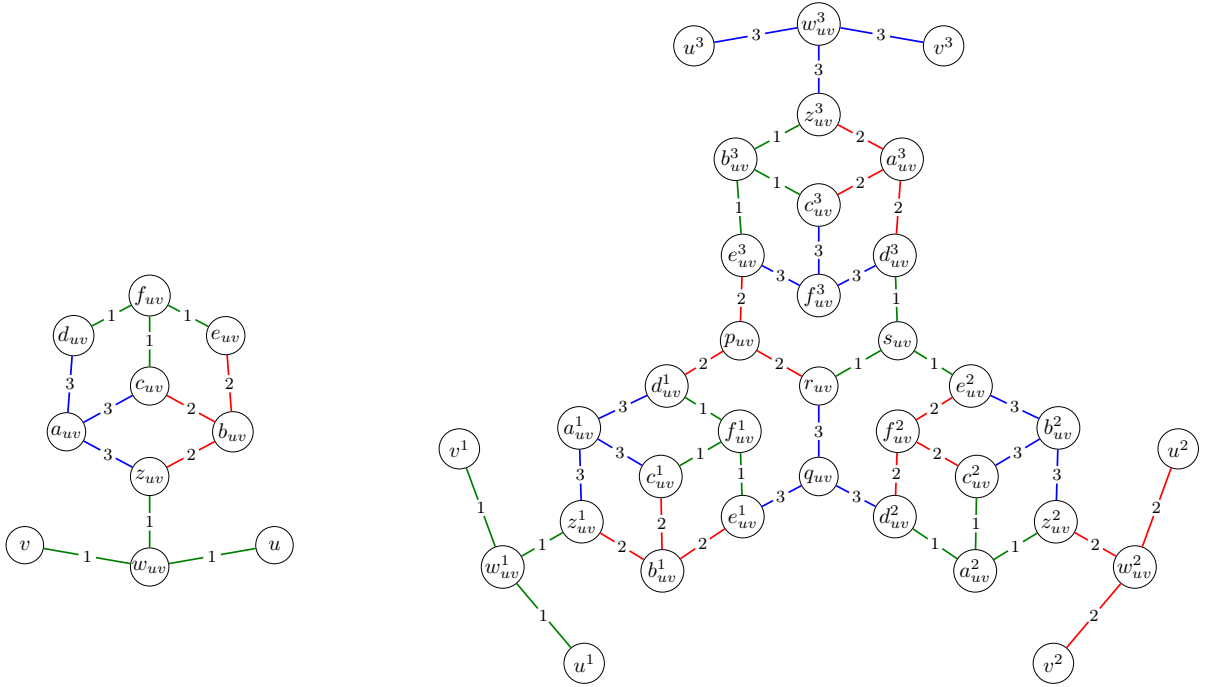

If $G$ is properly $3$-edge-colorable, then we fix such a coloring $\gamma: E(G)\to\set{1,2,3}$. For $i \in \set{1,2,3}$, we color the edges incident with $w_{uv}^i$ with the color $\gamma(uv) + i$, where the colors are considered to be taken modulo $3$ (considering $0=3$). Then it suffices to extend the obtained coloring to each edge gadget (see Figure~\ref{fig:cubic-3colors-fig}).
\end{proof}

\subsection{Proof of Theorem~\ref{th:cubic}.\ref{th:4-cubic}}

\begin{proof}
Let $G$ be the input graph. For each vertex $u$ of $G$, we replace it by the following vertex gadget $S_u$ (see Figure~\ref{fig:4colors}). The gadget $S_u$ is made of a 9-cycle $x_0^ux_1^u\dots x_8^u$ and three other vertices $y_i^u$ ($i \in \set{0,3,6}$) that will be connected to the rest of the graph. We add the edges $x_1^ux_8^u$, $x_2^ux_4^u$,  $x_5^ux_7^u$, $x_0^uy_0^u$, $x_3^uy_3^u$ and $x_6^uy_6^u$. For any edge-coloring $\gamma$ of $S_u$, we note $C_i^u(\gamma) = \set{\gamma(x_i^ux_{i+1}^u), \gamma(x_i^ux_{i-1}^u)}$ where $i \in \set{0,3,6}$ and where the indices are taken modulo~$9$. 

\begin{claim}\label{cl:4-vertex}
For every injective $4$-edge-coloring $\gamma$ of $S_u$ and for  every $i \in \set{0,3,6}$, the color $\gamma(x_i^uy_i^u)$ belongs to the set $C_i^u(\gamma)$. Moreover, $C_0^u(\gamma) \cup C_3^u(\gamma) \cup C_6^u(\gamma) = \set{1,2,3,4}$ and there exists a color $a \in \set{1,2,3,4}$ such that for all $i \in \set{0,3,6}$, $ a \in C_i^u(\gamma)$.

Furthermore, for any choice of color for $x_0^uy_0^u$, $x_3^uy_3^u$, $x_6^uy_6^u$ and sets of colors $C_i^u(\gamma)$, $i \in \set{0,3,6}$ verifying the previous necessary conditions, there exists an injective $4$-edge-coloring $\gamma$ of $S_u$ matching those choices.
\end{claim}

\begin{subproof}
Let us try to construct an injective $4$-edge-coloring $\gamma$ of $S_u$. Up to permuting the colors, we assume that $\gamma(x_0^ux_1^u)=1$, $\gamma(x_0^ux_8^u)=2$ and $\gamma(x_8^ux_1^u)=3$. Note that $x_2^ux_4^u$ and $x_5^ux_7^u$ cannot both be colored~$4$, w.l.o.g. assume that $\gamma(x_2^ux_4^u)\neq 4$. Hence $\gamma(x_2^ux_4^u)=2$ and $\gamma(x_2^ux_3^u)=4$. Remark that $\gamma(x_5^ux_6^u)\neq 2$. Moreover  $x_5^ux_7^u$ and  $x_6^ux_7^u$ can only receive colors $1$ or $4$ and they must receive different colors. Hence $\gamma(x_5^ux_6^u)=3$, $\gamma(x_3^ux_4^u)=1$, $\gamma(x_5^ux_7^u)=4$ and $\gamma(x_6^ux_7^u)=1$. 
Now there are two ways to complete the coloring of $S_u$, either $\gamma(x_1^ux_2^u)=4$, $\gamma(x_4^ux_5^u)=3$ and $\gamma(x_7^ux_8^u)=2$ or, $\gamma(x_1^ux_2^u)=3$, $\gamma(x_4^ux_5^u)=2$ and $\gamma(x_7^ux_8^u)=4$. In both cases all properties of the first part of the claim hold (with $a=1$).

Finally, note that the second of the two previous coloring options allows us to color $x_i^uy_i^u$, $i \in \set{0,3,6}$ with any color among those of $x_i^ux_{i+1}^u$ and $x_i^ux_{i-1}^u$, and to complete the coloring.
\end{subproof}

\begin{figure*}
    \centering
    \scalebox{.8}{\begin{tikzpicture}
\tikzstyle{n} = [draw, circle, inner sep=0pt, text width=6mm, align=center];

\foreach \i in {0,1,...,8}{
    \node[n] (x\i) at (\i*360/9:2) {$x_{\i}^u$};
    \node[n] (z\i) at ($(9,0)+ (180+\i*360/9:2)$) {$x_{\i}^v$};
}

\foreach \i in {0,3,6}{
    \node[n] (y\i) at (\i*360/9:3.5) {$y_{\i}^u$};
    \node[n] (w\i) at ($(9,0)+ (180+\i*360/9:3.5)$) {$y_{\i}^v$};
    \draw[thick] (x\i) -- (y\i);
    \draw[thick] (z\i) -- (w\i);
}

\draw[thick] (x0) -- (x1) -- (x2) -- (x3) -- (x4) -- (x5) -- (x6) -- (x7) -- (x8) -- (x0);
\draw[thick] (z0) -- (z1) -- (z2) -- (z3) -- (z4) -- (z5) -- (z6) -- (z7) -- (z8) -- (z0);

\draw[thick] (x1) -- (x8) (x2) -- (x4) (x5) -- (x7) (z1) -- (z8) (z2) -- (z4) (z5) -- (z7);

\node[n] (w) at (4.5,1) {$w_{uv}$};
\node[n] (z) at (4.5,-1) {$z_{uv}$};

\draw[thick] (y0) -- (w) -- (w0) -- (z) -- (y0);
\draw[thick] (w)  -- (z);

%\myedge[\EdgeUn]{(u1)}{(w1)}{1}

\end{tikzpicture}}
    \caption{Two vertex gadgets $S_u$ and $S_v$, corresponding to the vertices $u$ and $v$ of a graph $G$, connected by an edge gadget corresponding to the edge $uv$ of $G$.}
    \label{fig:4colors}
\end{figure*}
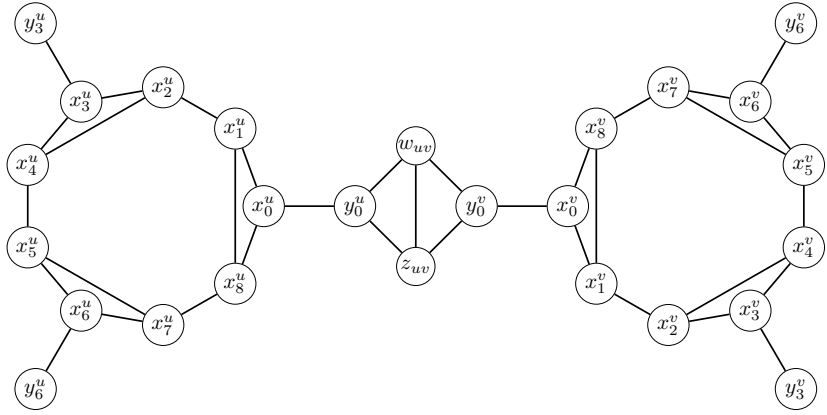

For every edge $uv$ of $G$, we construct the following edge gadget $E_{uv}$ (see Figure~\ref{fig:4colors}). First, choose $y_i^u$ (resp. $y_j^v$) of degree~1 among the vertices of $S_u$ (resp. $S_v$). Create two new adjacent vertices $w_{uv}$ and $z_{uv}$ such that $y_i^uw_{uv}y_j^vz_{uv}$ is a $4$-cycle.

\begin{claim}\label{cl:4-edge}
For every injective $4$-edge-coloring $\gamma$ of $G$ and every edge gadget $E_{uv}$ connecting $y_i^u$ and $y_j^v$ ($i,j \in \set{0,3,6})$, we have $C_i^u(\gamma) = C_j^v(\gamma)$.

Furthermore, any injective $4$-edge-coloring $\gamma$ of $S_u$ and $S_v$ such that $C_i^u(\gamma) = C_j^v(\gamma)$ and $\gamma(x_i^uy_{i}^u)= \gamma(x_j^vy_{j}^v)$ can be extended to an injective $4$-edge-coloring of $S_u \cup E_{uv} \cup S_v$.
\end{claim}

\begin{subproof}
Suppose, w.l.o.g. by Claim~\ref{cl:4-vertex}, that $x_i^ux_{i+1}^u$ is colored $1$, $x_i^ux_{i-1}^u$ is colored $2$ and $x_i^uy_{i}^u$ is colored $1$. Now w.l.o.g., $y_i^uw_{uv}$ is colored $3$ and  $y_i^uz_{uv}$ is colored $4$. This implies that $w_{uv}z_{uv}$ is colored $2$, $y_j^vw_{uv}$ is colored $3$, $y_j^vz_{uv}$ is colored $4$, $y_j^vx_j^v$ is colored $1$ and $C_j^v(\gamma) = \set{1,2}$.

The second part of the claim is proved by taking the previous coloring and extending it using the second part of Claim~\ref{cl:4-vertex}.
\end{subproof}

Let $G'$ be the cubic graph constructed from $G$ by the above process. By Claim~\ref{cl:4-edge}, if $uv$ is an edge connecting $y_i^u$ and $y_j^v$ then for any injective coloring $\gamma$ of $G'$, $C_i^u(\gamma) = C_j^v(\gamma) = \set{a,b}$ for some $a$ and $b$. Hence this set somehow characterizes the edge gadget $E_{uv}$, we say that $E_{uv}$ is {\em colored} by $\set{a,b}$. 

Suppose that there exists an injective $4$-edge-coloring $\gamma$ of $G'$. For each edge $uv$ of $G$, we color  $uv$  depending on the coloring of $E_{uv}$. When $E_{uv}$ is colored $\set{1,2}$ or $\set{3,4}$ (resp. $\set{1,3}$ or $\set{2,4}$, resp. $\set{1,4}$ or $\set{2,3}$) then we color $uv$ by color $1$ (resp. $2$, resp. $3$). We argue that this edge-coloring, noted $\gamma$, is proper. Indeed suppose it is not, then for some vertex $u$, w.l.o.g., $uv$ and $uw$ are both colored~$1$. This means that the coloring of $G'$ is such that $C_i^u(\gamma) = C_j^u(\gamma)$ or $C_i^u(\gamma) \cap C_j^u(\gamma) = \varnothing$ for $i \neq j$ and $i,j \in \set{0,3,6}$. This contradicts Claim~\ref{cl:4-vertex}. Hence we get a proper $3$-edge-coloring of $G$.

Conversely, suppose that there exists a proper $3$-edge-coloring of $G$. In $G'$, we color each edge of the form $x_i^uy_i^u$ by $1$. If an edge $uv$ of $G$ is colored $1$ (resp. $2$, resp. $3$) then we assign the color $\set{1,2}$ (resp. $\set{1,3}$, resp. $\set{1,4}$) to $E_{uv}$. By Claim~\ref{cl:4-vertex}, this coloring can be extended to an injective $4$-edge-coloring of each $S_u$, $u \in V(G)$. By Claim~\ref{cl:4-edge}, this injective $4$-edge-coloring can be extended to each edge gadget to color the whole graph.
%Hence $G$ admits a proper $3$-edge-coloring if and only if $G'$ admits an injective $4$-edge-coloring.
\end{proof}

\section{Proof of Theorem~\ref{th:3-planar}}

We will reduce from the following problem:

\medskip\noindent
\textsc{Planar $3$-Vertex-Coloring} \\
Instance: A planar graph $G$ with maximum degree~$4$.\\
Question: Does $G$ admit a proper $3$-vertex-coloring?
\medskip

This problem was proven to be NP-Complete in \cite{GJS76}.
Let $G$ be a planar graph with maximum degree~$4$.

\subsection{Proof of Theorem~\ref{th:3-planar}.1}

\begin{proof}
Recall that we want to construct a graph $G'$ with girth at least $g$.

For each vertex $u \in V(G)$, we construct a vertex gadget $S_u$ as follows (see Figure~\ref{fig:cubic-planar-3colors-fig}).
First create a cycle $x^u_1x^u_2\dots x^u_\ell$ where $\ell \geq g$ and $\ell$ is an odd multiple of $3$. To each $x^u_i$ add a single pendant neighbor $y^u_i$ of degree $1$. To the vertex $y^u_1$, add two non-adjacent neighbors $w^u$ and $z^u$. Create four more vertices $a^u_1$, $b^u_1$, $c^u_1$ and $d^u_1$. 
The vertex $w^u$ is adjacent to $a^u_1$ and $b^u_1$ while $z^u$ is adjacent to  $c^u_1$ and $d^u_1$. Now construct a path $a^u_1a^u_2\dots a^u_g$ of length $g$ and add to each $a^u_i$ for $i \leq g-1$ a pendant vertex of degree $1$ called $a'^u_i$. 
Similarly we create the vertices $b^u_1\dots b^u_g, b'^u_1 \dots b'^u_{g-1}$, $c^u_1\dots c^u_g, c'^u_1 \dots c'^u_{g-1}$ and $d^u_1\dots d^u_g, d'^u_1 \dots d'^u_{g-1}$.
Finally add a vertex $\alpha^u$ (resp. $\beta^u$, resp. $\gamma^u$, resp. $\delta^u$) adjacent to $a^u_g$ (resp. $b^u_g$, resp. $c^u_g$, resp. $d^u_g$).

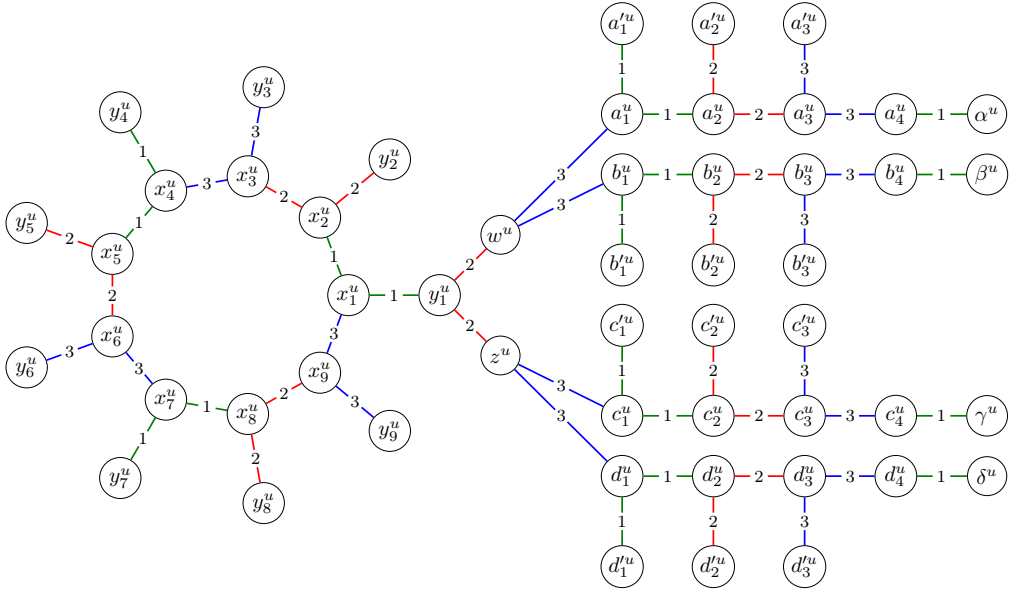
\begin{figure*}[th]
    \centering
    \scalebox{.8}{\begin{tikzpicture}
\tikzstyle{n} = [draw, circle, inner sep=0pt, text width=6mm, align=center];
%\node[n] (r) at (0,0) {$r_{uv}$};

%\node[n] (s) at (0:1.5) {$s_{uv}$};
%\node[n] (p) at (360/3:1.5) {$p_{uv}$};
%\node[n] (q) at (2*360/3:1.5) {$q_{uv}$};

\foreach \i in {1,...,9}{
    \node[n] (x\i) at (\i*360/9-360/9:2) {$x_{\i}^{u}$}; 
    \node[n] (y\i) at (\i*360/9-360/9:3.5) {$y_{\i}^{u}$}; 
    %\node[n] (d\i) at (\i*360/3 +360/6-360/12:2.5) {$d_{uv}^{\i}$}; 
    %\node[n] (c\i) at (\i*360/3 +360/6:3) {$c_{uv}^{\i}$}; 
    %\node[n] (b\i) at (\i*360/3 +360/6+360/18:4) {$b_{uv}^{\i}$}; 
    %\node[n] (a\i) at (\i*360/3 +360/6-360/18:4) {$a_{uv}^{\i}$}; 
    %\node[n] (z\i) at (\i*360/3 +360/6:4.5) {$z_{uv}^{\i}$}; 
    %\node[n] (w\i) at (\i*360/3 +360/6:6) {$w_{uv}^{\i}$}; 
    %\node[n] (u\i) at (\i*360/3 +360/6+360/18:6) {$u^{\i}$}; 
    %\node[n] (v\i) at (\i*360/3 +360/6-360/18:6) {$v^{\i}$}; 
}

\myedge[\EdgeUn]{(x1)}{(x2)}{1}
\myedge[\EdgeUn]{(x1)}{(y1)}{1}

\myedge[\EdgeDeux]{(x2)}{(x3)}{2}
\myedge[\EdgeDeux]{(x2)}{(y2)}{2}

\myedge[\EdgeTrois]{(x3)}{(x4)}{3}
\myedge[\EdgeTrois]{(x3)}{(y3)}{3}

\myedge[\EdgeUn]{(x4)}{(x5)}{1}
\myedge[\EdgeUn]{(x4)}{(y4)}{1}

\myedge[\EdgeDeux]{(x5)}{(x6)}{2}
\myedge[\EdgeDeux]{(x5)}{(y5)}{2}

\myedge[\EdgeTrois]{(x6)}{(x7)}{3}
\myedge[\EdgeTrois]{(x6)}{(y6)}{3}

\myedge[\EdgeUn]{(x7)}{(x8)}{1}
\myedge[\EdgeUn]{(x7)}{(y7)}{1}

\myedge[\EdgeDeux]{(x8)}{(x9)}{2}
\myedge[\EdgeDeux]{(x8)}{(y8)}{2}

\myedge[\EdgeTrois]{(x9)}{(x1)}{3}
\myedge[\EdgeTrois]{(x9)}{(y9)}{3}

\node[n] (w) at (4.5,1) {$w^u$};
\node[n] (z) at (4.5,-1) {$z^u$};

\foreach \l/\c in {a/3, b/2, c/-2, d/-3}{
    \foreach \i in {1, ...,4}{
        \node[n] (\l\i) at (5+1.5*\i,\c) {$\l_\i^u$};
    }
}

\node[n] (alpha) at (5+1.5*5,3) {$\alpha^u$};
\node[n] (beta) at (5+1.5*5,2) {$\beta^u$};
\node[n] (gamma) at (5+1.5*5,-2) {$\gamma^u$};
\node[n] (delta) at (5+1.5*5,-3) {$\delta^u$};

\foreach \l/\c in {a/4.5, b/0.5, c/-0.5, d/-4.5}{
    \foreach \i in {1, ...,3}{
        \node[n] (\l\i') at (5+1.5*\i,\c) {$\l_\i'^u$};
    }
}

\Deux{(y1)}{(w)}
\Deux{(y1)}{(z)}
\Trois{(w)}{(a1)}
\Trois{(w)}{(b1)}
\Trois{(z)}{(c1)}
\Trois{(z)}{(d1)}

\foreach \l in {a, b, c, d}{
   \Un{(\l1)}{(\l2)}
   \Un{(\l1)}{(\l1')}
   
   \Deux{(\l2)}{(\l3)}
   \Deux{(\l2)}{(\l2')}
   
   \Trois{(\l3)}{(\l4)}
   \Trois{(\l3)}{(\l3')}
}
\Un{(a4)}{(alpha)}
\Un{(b4)}{(beta)}
\Un{(c4)}{(gamma)}
\Un{(d4)}{(delta)}
\end{tikzpicture}}
    \caption{Vertex gadget $S_u$ for planar subcubic graphs with girth at least $g$ (in this example $g=4$ and $\ell = 9$).}
    \label{fig:cubic-planar-3colors-fig}
\end{figure*}

\begin{claim}\label{cl:planar_big_girth}
For any injective $3$-edge-coloring $\rho$ of $S_u$, $\rho(a^u_g\alpha^u) = \rho(b^u_g\beta^u)= \rho(c^u_g\gamma^u) = \rho(d^u_g\delta^u)$. We call this color $\rho(S_u)$. Moreover, for any choice of a color $\rho(S_u)$, there exists an injective $3$-edge-coloring $\rho$ with these properties.
\end{claim}
\begin{subproof}
%Take an injective $3$-edge-coloring $\rho$ of $S_u$. 
Suppose that there exists $i \in \set{1,\dots,\ell}$ such that the property $\mathcal P(i) := ``\rho(x^u_ix^u_{i+1}) = \rho(x^u_iy^u_{i}) \neq \rho(x^u_ix^u_{i-1})$'' holds (the indices are taken modulo $\ell$, considering $0=\ell$). Then $\mathcal P(i)$ holds for all $i \in \set{1,\ldots,\ell}$. Indeed, take such an $i$, then $\rho(x^u_{i+1}x^u_{i+2}) = \rho(x^u_{i+1}y^u_{i+1})$ is the color $\set{1,2,3} \setminus \set{\rho(x^u_iy^u_{i}), \rho(x^u_ix^u_{i-1})}$. Hence the property holds for $i+1$, by induction it holds for every $i$.
Note that the same can be said for the property $\mathcal P'(i) = ``\rho(x^u_ix^u_{i-1}) = \rho(x^u_iy^u_{i}) \neq \rho(x^u_ix^u_{i+1})$''. Also note that if $\rho(x^u_ix^u_{i-1}) =  \rho(x^u_ix^u_{i+1})  \neq \rho(x^u_iy^u_{i})$ then we have $\mathcal P(i+1)$ which is a contradiction because we do not have $\mathcal P(i)$.

Suppose now that for all $i$, neither $\mathcal P(i)$ nor $\mathcal P'(i)$ holds. This means that the edges incident to a vertex $x^u_i$ are either of the same color, or of three distinct colors. If they have the same color, then the edges incident with $x^u_{i+1}$ have three distinct colors, the ones incident to $x^u_{i+2}$ have the same color, and so on. This would imply that the cycle $x^u_1\dots x^u_{\ell}$ is even, which is a contradiction. Moreover, if the edges incident to $x^u_i$ have three distinct colors, then the edges incident to $x^u_{i+1}$ (or $x^u_{i-1}$) would all have the same color,
and therefore no injective 3-edge-coloring would be possible.

Thus, w.l.o.g. we can suppose that $\rho(x^u_1x^u_{2}) = \rho(x^u_1y^u_{1}) = 1$ and $\rho(x^u_1x^u_{\ell})=3$. By extending the coloring to the rest of $S_u$, we can infer that $\rho(y^u_1w^u)=\rho(y^u_1z^u)=2$, $\rho(w^ua^u_{1})=\rho(w^ub^u_{1})=3$ and $\rho(z^uc^u_{1})=\rho(z^ud^u_{1})=3$. By the same reasoning,  we can see that all the edges of $S_u$ (ignoring the edges involving one of the vertices $x^u_i$) have only one possible color which depends only on their distance to $y^u_1$ and in particular $\rho(a^u_g\alpha^u) = \rho(b^u_g\beta^u)= \rho(c^u_g\gamma^u) = \rho(d^u_g\delta^u)$.

%Conversely, by extending this coloring, we obtain an injective $3$-edge-coloring of all the edges of $S_u$ except those incident to a vertex $x^u_i$. For these edges, we can also extend the coloring: $\rho(x^u_iy^u_{i}) =\rho(x^u_ix^u_{i+1}) = i \pmod 3$ (we consider that color $0$ is also color $3$) and $\rho(x^u_ix^u_{i-1}) = i-1 \pmod 3$. Since the number of vertices $x^u_i$ is a multiple of $3$, there is no problem with this coloring. Hence, we obtain a coloring of $S_u$. To choose the color $\rho(S_u)$, it suffices to permute the colors in the previous coloring.
Conversely, $S_u$ admits a coloring (see Figure~\ref{fig:cubic-planar-3colors-fig} for an example). To choose a coloring of $S_u$ having the desired color $\rho(S_u)$, it suffices to permute the colors in the previous coloring.
\end{subproof}

To finish the construction, for any edge $uv \in E(G)$, we add an edge $e^{uv}$ to $G'$ between a vertex among $\set{\alpha^u,\beta^u,\gamma^u, \delta^u}$ and a vertex among $\set{\alpha^v,\beta^v,\gamma^v, \delta^v}$ such that the planarity of $G'$ is preserved. This can be done by cyclically ordering the vertices of $\set{\alpha^u,\beta^u,\gamma^u, \delta^u}$ according to a planar embedding of $G$, and adding the edge $e ^{uv}$ between the right pair of vertices.

Note that $G'$ is planar, subcubic with girth at least $g$.

\medskip
Suppose that $G'$ admits an injective $3$-edge-coloring $\rho$. Assign to the vertex $u$ of $G$ the color $\rho(S_u)$. 
%This coloring of $G$ is proper. Indeed, if it was not the case that there would be some adjacent vertices $u$ and $v$ receiving the same color. 
Take two adjacent vertices $u$ and $v$ of $G$.
The edge $e^{uv}$ in $G'$ is an edge between two vertices, one of $S_u$ and one of $S_v$: w.l.o.g. say $e^{uv} = \alpha^u\alpha^v$. This implies that $a^u_g\alpha^u$ and $a^v_g\alpha^v$ receive different colors and thus $\rho(S_u) \neq \rho(S_v)$. Hence this coloring of $G$ is a proper $3$-vertex-coloring.

\medskip
Conversely, suppose that $G$ admits a proper $3$-vertex-coloring. Let $\rho$ be a partial edge-coloring of $G'$ with no colored edges. We choose the color $\rho(S_u)$ to be the color of $u$ in $G$ (and we color the appropriate edges of $G'$). By Claim~\ref{cl:planar_big_girth}, we can extend $\rho$ to each gadget $S_u$. Note that by the choice of $\rho(S_u)$, there is no conflict between edges of $S_u$ and $S_v$ when $u$ and $v$ are adjacent in $G$. It is left to color the edges of the form $e^{uv}$. By construction, there are only two edges at distance~$2$ of $e^{uv}$ (and this edge does not belong to a triangle). Hence there is at least one remaining color for $e^{uv}$. After coloring theses edges, $\rho$ is an injective $3$-edge-coloring of $G'$.
\end{proof}

\subsection{Proof of Theorem~\ref{th:3-planar}.2}

\begin{proof}
In order to prove this result, we will modify the previous construction to make it bipartite (the girth condition will be lost).

\begin{figure*}
    \centering
    \scalebox{.8}{\begin{tikzpicture}[scale=0.75]
\tikzstyle{n} = [draw, circle, inner sep=0pt, text width=7mm, align=center];
\node[n] (x) at (0,0) {$x_{3,1}^u$};

%\node[n] (s) at (0:1.5) {$s_{uv}$};
%\node[n] (p) at (360/3:1.5) {$p_{uv}$};
%\node[n] (q) at (2*360/3:1.5) {$q_{uv}$};

%\foreach \i in {1,...,6}{
%    \node[n] (x\i) at (\i*360/6:4) {}; 
%}

\node[n] (x1) at (60:4) {$x_{1,1}^u$};
\node[n] (x2) at (120:4) {$x_{14,1}^u$};
\node[n] (x3) at (180:4) {$x_{4,1}^u$};
\node[n] (x4) at (240:4) {$x_{24,1}^u$};
\node[n] (x5) at (300:4) {$x_{2,1}^u$};
\node[n] (x6) at (0:4) {$x_{12,1}^u$};

\node[n] (y1) at (60:2) {$x_{13,1}^u$};
\node[n] (y3) at (180:2) {$x_{34,1}^u$};
\node[n] (y5) at (300:2) {$x_{23,1}^u$};

\node[n] (w1) at (120:2) {$y_{13,1}^u$};
\node[n] (w3) at (240:2) {$y_{34,1}^u$};
\node[n] (w5) at (0:2) {$y_{23,1}^u$};

\node[n] (z2) at (120:6) {$y_{14,1}^u$};
\node[n] (z4) at (240:6) {$y_{24,1}^u$};

\node[n] (x7) at (0:6) {$y_{12,1}^u$};
\node[n] (x8) at (0:8) {$y_{12,2}^u$};

\Un{(x1)}{(x6)}
\Un{(x5)}{(x6)}
\Un{(x7)}{(x6)}

\Un{(y3)}{(x3)}
\Un{(y3)}{(w3)}
\Un{(y3)}{(x)}

\Deux{(x2)}{(x1)}
\Deux{(x2)}{(x3)}
\Deux{(x2)}{(z2)}

\Deux{(y5)}{(x5)}
\Deux{(y5)}{(w5)}
\Deux{(y5)}{(x)}

\Trois{(x4)}{(x5)}
\Trois{(x4)}{(x3)}
\Trois{(x4)}{(z4)}

\Trois{(y1)}{(x1)}
\Trois{(y1)}{(w1)}
\Trois{(y1)}{(x)}

\Trois{(x7)}{(x8)}

%%%%%changer l originie du repère en y pour faire un copier-coller de tout ce qui se trouve au-dessus

\node[n] (y) at ($(x) + (0:14)$) {$x_{3,2}^u$};

\node[n] (x'1) at ($(y) - (60:4)$) {$x_{1,2}^u$};
\node[n] (x'2) at ($(y) - (120:4)$) {$x_{14,2}^u$};
\node[n] (x'3) at ($(y) - (180:4)$) {$x_{4,2}^u$};
\node[n] (x'4) at ($(y) - (240:4)$) {$x_{24,2}^u$};
\node[n] (x'5) at ($(y) - (300:4)$) {$x_{2,2}^u$};
\node[n] (x'6) at ($(y) - (0:4)$) {$x_{12,2}^u$};

\node[n] (y'1) at ($(y) - (60:2)$) {$x_{13,2}^u$};
\node[n] (y'3) at ($(y) - (180:2)$) {$x_{34,2}^u$};
\node[n] (y'5) at ($(y) - (300:2)$) {$x_{23,2}^u$};

\node[n] (w'1) at ($(y) - (120:2)$) {$y_{13,2}^u$};
\node[n] (w'3) at ($(y) - (240:2)$) {$y_{34,2}^u$};
\node[n] (w'5) at ($(y) - (0:2)$) {$y_{23,2}^u$};

\node[n] (z'2) at ($(y) - (120:6)$) {$y_{14,2}^u$};
\node[n] (z'4) at ($(y) - (240:6)$) {$y_{24,2}^u$};
\node[n] (x'7) at ($(x7) + (270:2)$) {$y_1^u$};

\node[n] (y'7) at ($(x'7) + (240:2)$) {$w^u$};
\node[n] (y'8) at ($(x'7) + (300:2)$) {$z^u$};
\node (dots) at ($(x'7) + (-90:2.5)$) {$\vdots$};

\Deux{(x'1)}{(x'6)}
\Deux{(x'5)}{(x'6)}
\Deux{(x8)}{(x'6)}

\Deux{(y'3)}{(x'3)}
\Deux{(y'3)}{(w'3)}
\Deux{(y'3)}{(y)}

\Trois{(x'2)}{(x'1)}
\Trois{(x'2)}{(x'3)}
\Trois{(x'2)}{(z'2)}

\Trois{(y'5)}{(x'5)}
\Trois{(y'5)}{(w'5)}
\Trois{(y'5)}{(y)}

\Un{(x'4)}{(x'5)}
\Un{(x'4)}{(x'3)}
\Un{(x'4)}{(z'4)}

\Un{(y'1)}{(x'1)}
\Un{(y'1)}{(w'1)}
\Un{(y'1)}{(y)}

\Trois{(x7)}{(x'7)}
\Deux{(x'7)}{(y'7)}
\Deux{(x'7)}{(y'8)}

\end{tikzpicture}}
    \caption{Vertex gadget for planar bipartite subcubic graphs with girth at least $6$.}
    \label{fig:planar-bipartite-3colors-fig}
\end{figure*}
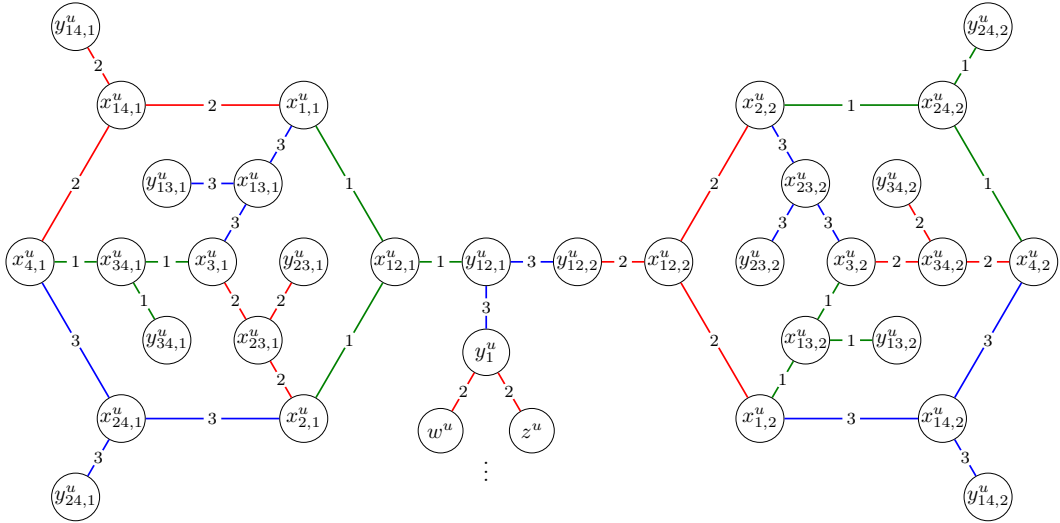

First we modify $S_u$ (see Figure~\ref{fig:planar-bipartite-3colors-fig}). 
Create the following gadget $H$. Start with a complete graph on four vertices $x_1, \ldots, x_4$. For each edge $x_ix_j$, create a vertex $x_{ij}$ adjacent to both $x_i$ and $x_j$ and remove the edge $x_ix_j$. To each of these vertices of degree~$2$, add a pendant edge, with $y_{ij}$ the vertex of degree~$1$ adjacent to $x_{ij}$.

We claim that in every injective $3$-edge-coloring $\gamma$ of $H$, for any $i \neq j$, the vertex $x_{ij}$ is incident to only one color. Suppose it is not the case, then there must exist an injective $3$-edge-coloring $\gamma$ for which we have one of $x_{12}x_2$ and $x_{12}x_{1}$ colored differently from $x_{12}y_{12}$, say w.l.o.g. $\gamma(x_{12}x_{1}) =1$  and $\gamma(x_{12}y_{12}) = 2$. We deduce that $\gamma(x_{2}x_{23}) =\gamma(x_{2}x_{24}) =3$, $\gamma(x_{14}x_{4})=\gamma(x_{3}x_{13})=2$, $\gamma(x_{3}x_{34}) =1$, and there is no color available for $x_{23}y_{23}$, a contradiction. %It is easy to check that the case where $\gamma(x_{12}x_1)=1$, $\gamma(x_{12}x_2)=2$ is similar.
%$\gamma(x_{31}x_{1}) =3$, $\gamma(x_{14}y_{14}) =2$ and there is no color available for $x_4x_{24}$, a contradiction.

Now, take two disjoint copies of $H$ named $H^u_1$ and $H^u_2$. Add an edge between the two vertices $y_{12,1}^u$ and $y_{12,2}^u$ and add the edge $y_{12,1}^uy^u_1$ where $y^u_1$ is a new vertex. Now repeat the construction process of $S_u$, for $g=6$ for example, as described in the previous section by starting at the step where the vertices $w^u$ and $z^u$ are added. As we observed, the edges incident to vertex $x_{12,1}^u$ of $H_1^u$ (resp. $x_{12,2}^u$ of $H_2^u$) have the same color in any injective $3$-edge-coloring $\rho$.
Hence, $\rho(y_{12,1}^uy_{12,2}^u) = \rho(y_{12,1}^uy^u_1) \neq \rho(x_{12,1}^uy_{12,1}^u)$. Note that this graph also admits an injective $3$-edge-coloring (see Figure~\ref{fig:planar-bipartite-3colors-fig}). We are in the same configuration as in the proof of Theorem~\ref{th:4-cubic}.1. Thus Claim~\ref{cl:planar_big_girth} also holds for this gadget $S_u$.
Note that this gadget is bipartite.

The edge gadget does not change, it is still the edge $e^{uv}$. We need to be careful with the bipartiteness of the constructed graph. To ensure that the constructed graph is bipartite, it suffices that all vertices $y_1^u$, $u\in V(G)$, belong to the same part of the bipartition. To that end, if there is a path of odd length between $y_1^u$ and $y_1^v$, then w.l.o.g. this path is $y_1^u a^u_1\dots a^u_g \alpha^u \alpha^v a^v_g\dots a^v_1 y_1^v$. If we increase the length of a sequence $a^u_1\dots a^u_g$ in $S_u$ by~$3$ (and also adding $a'^u_{g}$, $ a'^u_{g+1}$ and $a'^u_{g+2}$), then this path now has even length. With this trick, we can ensure the bipartiteness of the constructed graph $G'$ as well as keeping Claim~\ref{cl:planar_big_girth} true in this new setting. 

Hence, as before, $G$ admits a proper vertex-$3$-coloring if and only if $G'$ admits an injective $3$-edge-coloring.\end{proof}

\section{Proof of Theorem~\ref{th:maille16}}

\begin{proof}
Let $G$ be a planar bipartite subcubic graph with girth at least $16$. Let $A$ and $B$ be the two parts of the bipartition of $G$. We construct the graph $G_A$ as follows: for each $u \in A$, we create a vertex $u$ in $G_A$. For each pair of vertices $u$, $v$ of $A$ which are at distance $2$, we add an edge between $u$ and $v$ in $G_A$.
As $G$ is subcubic, a planar embedding of $G$ also serves as a planar embedding of $G_A$, where the edges of $G_A$ follow their corresponding path of length~$2$ in $G$.
Hence, $G_A$ is a planar graph with maximum degree at most~$6$. Note that, by the girth condition on $G$, $G_A$ does not have any $k$-cycle, for all $k$ with $4\leq k\leq 7$. 
Then, by the main result from~\cite{Borodin05}, the graph $G_A$ admits a vertex-$3$-coloring $\gamma$. 

We now color $G$ as follows: each edge $uv$ of $G$, where $u\in A$ and $v\in B$, is colored by the color $\gamma(u)$ in $G_A$. We claim that this is an injective $3$-edge-coloring of $G$. Indeed, take any path $uvwz$ of $G$. W.l.o.g., assume $u,w \in A$ and $v,z \in B$. By construction, $uw \in E(G_A)$ and thus $uv$ and $wz$ receive different colors.
\end{proof}

\section{Proof of Theorem~\ref{th:fpt}}

\begin{proof}
We give an fixed-parameter tractable (FPT) algorithm parameterized by the treewidth $tw(G)$ of our input graph $G$. We use a \emph{nice tree decomposition} (see~\cite{niceTW}) of the input graph for our dynamic programming algorithm.
%Our algorithm is a dynamic programming algorithm using a \emph{nice tree decomposition} of the input graph. 
Nice tree decompositions are a well-known tool for designing  algorithms on graphs of bounded treewidth using dynamic programming. In our notation, the set of vertices of the graph associated to a node $v$ of the tree, its \emph{bag}, is denoted $X_v$. 

A nice tree decomposition of a  graph  is a tree decomposition, rooted at a node $Root$, with the following types of nodes. A \emph{join node} has exactly two children, with the same bags as their parent join node. An \emph{introduce node} has a unique child and contains exactly one more vertex in its bag than its child's bag. A \emph{forget node} also has a unique child, but the forget node's bag has exactly one less vertex than its child's bag. A \emph{leaf node} is a leaf of the tree and contains no vertices. 
We call $G_{\leq v}$ the subgraph of $G$ induced by the subtree of the decomposition rooted at $v$ and $G_{v}$ the subgraph of $G$ induced by $X_v$. We note $N_H(u)$ for the neighborhood of a vertex $u$ in a subgraph $H$ of $G$.

We define the following set associated with a node $v$:
$$ \mathcal T_v = \set{t_1 : X_v \rightarrow \mathcal P(\set{1,2,\dots,k})^2} \times \set{t_2 : E(G_v) \rightarrow \set{1,2,\dots,k}}, $$ where $\mathcal P(X)$ is the power set of $X$. 
For $T \in \mathcal T_v$ with $T=(t_1,t_2)$, to simplify notation, we note $T[u]$ for $t_1(u)$ when $u \in X_v$ and $T[e] = t_2(e)$ when $e \in E(G_v)$. For a vertex $u \in X_v$, we also note $A_u$ and $B_u$ the two sets such that $T[u] = t_1(u) = (A_u,B_u)$.

The set $\mathcal V_{al}(v)$  is the subset of $\mathcal T_v$ such that $T \in \mathcal V_{al}(v)$ if and only if there exists an injective $k$-edge-coloring $\gamma$ of $G_{\leq v}$ such that:
\begin{enumerate}
    \item for all $u \in X_v$, $A_u = \set{\gamma(uw), w \in V(G_{\leq v}) \setminus X_v}$, \textit{i.e.} $A_u$ is the set of colors of the edges of $G_{\leq v}$ (not in $G_v$) incident with $u$,
    \item for all $u \in X_v$, $B_u = \set{\gamma(zw), zw \in E(G_{\leq v}) \setminus E(G_v) \text{ and } z \in N_{G_{\leq v}}(u)}$, \textit{i.e.} $B_u$ is the set of colors of the edges of $G_{\leq v}$ (not in $G_v$) at distance $2$  of $u$ (or contained in a triangle containing $u$),
    \item for all $e \in E(G_v)$, $T[e]$ is the color $\gamma(e)$.
\end{enumerate}
In this case we say that $\gamma$ is \emph{associated with $T$}.
Note that for each injective $k$-edge-coloring of $G_{\leq v}$, there exists an associated $T \in \mathcal T_v$ and hence, $T \in \mathcal V_{al}(v)$. The set $\mathcal V_{al}(v)$ is thus the set of $T \in \mathcal T_v$ associated with an injective $k$-edge-coloring of $G_{\leq v}$.

Note that $\mathcal V_{al}(Root) \neq \emptyset$ if and only if there exists an injective $k$-edge-coloring of $G$. We will compute $\mathcal V_{al}(Root)$ with a dynamic programing algorithm. Also note that $\abs{\mathcal T_v} \leq 2^{O(k \cdot tw(G)^2)}$.

\medskip 

First suppose that $v$ is a leaf node. Then $\mathcal V_{al}(v) = \mathcal T_v = \set{(\emptyset, \emptyset)}$. 

\medskip 

Suppose that $v$ is a forget node where $v'$ is its child node such that $X_v \cup \set{a} = X_{v'}$. Let $T \in \mathcal T_v$, $T \in \mathcal V_{al}(v)$ if and only if there exists an associated coloring $\gamma$ of $G_{\leq v}$. This coloring $\gamma$ is also a coloring of $G_{\leq v'}$ and thus is associated to a $T' \in \mathcal V_{al}(v')$. In this case, since $T$ and $T'$ share the same coloring $\gamma$, we have the following constraints on $T$ and $T'$:
\begin{itemize}
    \item for all $e \in E(G_v)$, $T[e] = T'[e] = \gamma(e)$,
    \item for all $u \in X_v$ such that $au \in E(G_{v'})$, $A_u = A'_u \cup \set{T[au]}$ and $B_u = B'_u \cup \set{T[aw], w \in X_v \cap N_G(a), w \neq u}$ where $T[u] = (A_u,B_u)$ and $T'[u] = (A'_u, B'_u)$,
    \item for all $u \in X_v$ such that $au \notin E(G_{v'})$, $A_u = A'_u$ and $B_u = B'_u \cup \set{T[aw], w \in X_v \cap N_G(u) \cap N_G(a)}$ where $T[u] = (A_u,B_u)$ and $T'[u] = (A'_u, B'_u)$.
\end{itemize}
The last two constraints reflect the fact that $A_u$ and $B_u$ must be updated after the removal of $a$. The only new colors that can be added to these sets come from edges incident with $a$. There are multiple cases, depending on whether $u$ and $a$ are adjacent or not, determining which colors of edges need to be added to these sets.

Hence, for all $T \in \mathcal V_{al}(v)$, it suffices to check whether there exists a $T' \in \mathcal V_{al}(v')$ for which the previous conditions are verified. This can be done in time $2^{O(k \cdot tw(G)^2)}$, as $T$ is uniquely determined by $T'$ in the above constraints.

\medskip 
Suppose that $v$ is an introduce  node where $v'$ is its child node such that $X_v  = X_{v'} \cup \set{a}$. Let $T \in \mathcal T_v$, $T \in \mathcal V_{al}(v)$ if and only if there exists an associated coloring $\gamma$ of $G_{\leq v}$. This coloring $\gamma$ is also a coloring of $G_{\leq v'}$ and thus is associated to a $T' \in \mathcal V_{al}(v')$. In other words $T$ is associated to a coloring $\gamma$ obtained by extending a coloring $\gamma'$ associated to some $T' \in \mathcal V_{al}(v')$. Thus $T' \in \mathcal V_{al}(v')$, we have the following constraints on $T$ and $T'$, in order to ensure that $\gamma$ is the extension of $\gamma'$:
\begin{itemize}
    \item for all $e \in E(G_{v'})$, $T[e] = T'[e]$,
    \item for all $u \in X_{v'}$, $T[u] = T'[u]$,
    \item for $T[a] = (A_a,B_a)$, $A_a = \emptyset$ and $B_a = \bigcup_{u \in X_v, ua \in E(G_v)} A_u$,
    \item the coloring of $X_v$ is an injective $k$-edge-coloring,
    \item for all $ua \in E(G_v)$, $T[ua] \notin B_u \cup \bigcup_{u' \in X_v, u' \neq u, u'a \in E(G_v)} A_{u'}$.
\end{itemize}
The first two constraints correspond to the fact that $\gamma$ is an extension of $\gamma'$. As $a$ is a new vertex, $A_a = \emptyset$ and the only colors in $B_a$ can be obtained by edges incident with some vertex $u \in X_v$ itself adjacent to $a$, hence the third constraint.  The last two constraints correspond to the fact that the coloring of the new edges around $a$ cannot be in conflict with edges already colored. The fourth constraint checks that no such conflict arises in $X_v$ and the fifth constraint ensures that for each new edge $ua$ the color $T[ua]$ does not appear around an edge at distance~$2$ from $a$ or $u$. 
For each $T'$, there are at most $2^{tw(G)}$ possible candidates to be added to $\mathcal V_{al}(v)$. Hence $2^{O(k \cdot tw(G)^2)}$ time is sufficient to compute $\mathcal V_{al}(v)$ from $\mathcal V_{al}(v')$.

\medskip 
Suppose that $v$ is a join node where $v_1$ and $v_2$ are its children nodes such that $X_v  = X_{v_1} = X_{v_2}$. Let $T \in \mathcal T_v$, $T \in \mathcal V_{al}(v)$ if and only if there exists an associated coloring $\gamma$ of $G_{\leq v}$. As both $G_{\leq v_1}$ and $G_{\leq v_2}$ are subgraphs of $G_{\leq v}$, $\gamma$ is also a coloring of $G_{\leq v_i}$ ($i \in \set{1,2})$ and thus is associated to a $T_i \in \mathcal V_{al}(v_i)$. In this case, since $T$, $T_1$ and $T_2$ share the same coloring $\gamma$, we have the following constraints on  $T$, $T_1$ and $T_2$:
\begin{itemize}
    \item for all $e \in E(G_{v})$, $T[e] = T_1[e] = T_2[e]$,
    \item for all $u \in X_v$, $A_u = A^1_u \cup A^2_u$ and $B_u = B^1_u \cup B^2_u$ where $T_i[u] = (A^i_u,B^i_u) $ for $i \in \set{1,2}$,
    \item for all $uw \in E(G_v)$, $A_u \cap A_w = \emptyset$.
\end{itemize}
The last constraint corresponds to the fact that the coloring is an injective $k$-edge-coloring (\textit{i.e.} with no conflicts between the two subtrees). Given $T_1 \in \mathcal V_{al}(v_1)$ and $T_2 \in \mathcal V_{al}(v_2)$, $T$ is uniquely determined by the above constraints. Hence it suffices to try all the pairs of $T_1,T_2$ and when the obtained set $T$ verifies all conditions, we can add it to $\mathcal V_{al}(v)$. This can be done in time $(2^{O(k \cdot tw(G)^2)})^2 = 2^{O(k \cdot tw(G)^2)}$.
\end{proof}

\section{Proof of Theorem~\ref{th:small-Delta-NPC}}

\begin{proof}
We reduce from \textsc{$k$-Edge-Coloring}, proven to be NP-Complete even for $k$-regular graphs in \cite{Leven1983}.

\medskip\noindent
\textsc{$k$-Edge-Coloring} \\
Instance: A $k$-regular graph $G$.\\
Question: Does $G$ admit a proper $k$-edge-coloring?
\medskip

We choose $p$ to be the largest integer such that $k = \binom{p}{2} + r$ (and thus $r <p$) and  recall that $k \geq 45$. Moreover we set $\ell =2 p$.

Let $G$ be the input $k$-regular graph. For $uv \in E(G)$, we define the edge gadget $E_{uv}$ as follows (see Figure~\ref{fig:delta}). First create the following vertices $a^{uv}$, $b^{uv}$, $x^{uv}_1$, \dots, $x^{uv}_{p-3}$, $c^{uv}$, $d^{uv}$, $e^{uv}$, $y^{uv}_1$, \dots, $y^{uv}_{r}$, $s^{uv}_1$, \dots, $s^{uv}_{2\ell}$. The vertices $s^{uv}_i$ have degree~$1$ in $E_{uv}$ and will be connected to the rest of the graph. The vertices $\set{x^{uv}_1, \dots, x^{uv}_{p-3}, a^{uv}, b^{uv}, c^{uv}}$ form a clique; this is also the case for $\set{x^{uv}_1, \dots, x^{uv}_{p-3}, a^{uv}, b^{uv}, d^{uv}}$ and $\set{y^{uv}_1, \dots, y^{uv}_{r}, d^{uv}}$. The vertex $e^{uv}$ is adjacent to $c^{uv}$, $d^{uv}$, $x^{uv}_1$, $\dots$, $x^{uv}_{p-3}$, $s^{uv}_1$, \dots, $s^{uv}_{2\ell}$. In the case where $r = 0$, \textit{i.e.} $k = \binom{p}{2}$, we delete $d^{uv}$.

\begin{figure*}
    \centering
    \scalebox{.8}{\begin{tikzpicture}
\tikzstyle{n} = [draw, circle, inner sep=0pt, text width=8mm, align=center];

\node[n] (e) at (0,0) {$e^{uv}$};

\node[n] (x1) at (-1,2.5) {$x^{uv}_1$};
\node[n] (x2) at (0,2.5) {$x^{uv}_2$};
\node (xdots) at (1,2.5) {$\cdots$};
\node[n] (x3) at (2,2.5) {$x^{uv}_{p-3}$};

\node[n] (a) at (-0.5,5) {$a^{uv}$};
\node[n] (b) at (1.5,5) {$b^{uv}$};
\node[n] (c) at (4,3.5) {$c^{uv}$};
\node[n] (d) at (-3,3.5) {$d^{uv}$};

\node[n] (y1) at (-5,5) {$y^{uv}_1$};
\node[n] (y2) at (-5,4) {$y^{uv}_2$};
\node (ydots) at (-5,3.1) {$\vdots$};
\node[n] (yr) at (-5,2) {$y^{uv}_{r}$};

\node[n] (s1) at (190:2) {$s^{uv}_1$};
\node[n] (s2) at (230:2) {$s^{uv}_2$};
\node (sdots) at (260:2) {$\cdots$};
\node[n] (sl) at (290:2) {$s^{uv}_{\ell}$};
\node (sdots2) at (315:2) {$\cdots$};
\node[n] (s2l) at (340:2) {$s^{uv}_{2\ell}$};

%\draw[thick] (e) edge (s1) edge (s2) edge (sl) edge (s2l);
\Un{(e)}{(s1)}
\Un{(e)}{(s2)}
\Un{(e)}{(sl)}
\Un{(e)}{(s2l)}
\draw[thick] (e) edge (x1) edge (x2) edge (x3);
\draw[thick] (e) edge[bend right] (c) edge[bend left] (d);

%\draw[thick] (d) edge (y1) edge (y2) edge (yr);

\draw[thick] (d) edge (-1.75,5) edge (-1.75,4) edge (-1.75,2);
\draw[thick] (d) edge (-4.25,5) edge (-4.25,4) edge (-4.25,2);

\draw[thick] (c) edge (2.75,5) edge (2.75,4) edge (2.75,2);

\Un{(a)}{(b)}

% clique y
\draw[line width=2pt] (-5.75,5.75) -- (-4.25,5.75) -- (-4.25,1.25) -- (-5.75,1.25) -- cycle;

% clique principale
\draw[line width=2pt] (-1.75,5.75) -- (2.75,5.75) -- (2.75,1.75) -- (-1.75,1.75) -- cycle;

\end{tikzpicture}}
    \caption{The edge gadget $E_{uv}$ when $r > 0$. The vertices inside each of the two rectangle form a clique. 
    The vertex $c^{uv}$ is adjacent to every vertex inside the largest rectangle. The vertex $d^{uv}$ is adjacent to every vertex inside the two rectangles.
    %A vertex with edges which connect to one border of a rectangle is adjacent to every vertex inside the rectangle.
    }
    \label{fig:delta}
\end{figure*}
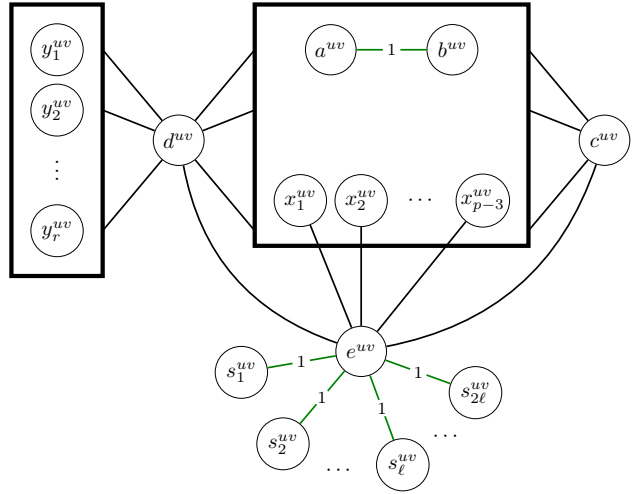

Let $u$ be a vertex of $G$ with $v_1, \dots, v_k$ its neighbors. We construct the vertex gadget $S_u$ from $k \times \ell$ vertices $v_{1,1}, \dots, v_{1,\ell}, v_{2,1}, \dots, v_{k,\ell}$ and successively consider pairs $v_i$, $v_j$ of neighbors. For each pair, we add an edge between one of $v_{i,1}, \dots, v_{i,\ell}$ of minimum degree and one of $v_{j,1}, \dots, v_{j,\ell}$ with minimum degree. By adding edges one by one in this way, we ensure that the maximum degree of the vertices of $S_u$ is at most $\frac{k}{\ell} +1$. %This can be done by ordering the edges and adding them one by one. When adding an edge, we choose iteratively joining vertices of smallest degree at each step.

Finally, for each edge $uv$ of $G$, we identify the $2\ell$ vertices   $s^{uv}_1$, \dots, $s^{uv}_{2\ell}$ with the $\ell$ vertices of $S_u$ corresponding to $v$ (since $v$ is a neighbour of $u$, by the construction of $S_u$ in the previous paragraph, there are $\ell$ such vertices in $S_u$) and with the $\ell$ vertices of $S_v$ corresponding to $u$. This creates the graph $G'$.
Note that its maximum degree is $\max(2\ell + p-1 ,\frac{k}{\ell} +2)  \leq 5p \leq 5\sqrt{3k}$.

\begin{claim}\label{cl:big-k:edge}
For any injective $k$-edge-coloring $\gamma$ of $E_{uv}$, we have $\gamma(e^{uv}s_1^{uv}) = \gamma(e^{uv}s_2^{uv}) = \dots =  \gamma(e^{uv}s_{2\ell}^{uv})$.
Moreover if $\gamma$ is a partial injective $k$-edge-coloring of $E_{uv}$ where $\gamma(e^{uv}s_1^{uv}) = \gamma(e^{uv}s_2^{uv}) = \dots = \gamma(e^{uv}s_{2\ell}^{uv})$ and there are no other colored edges, we can extend $\gamma$ to $E_{uv}$.
\end{claim}

\begin{subproof}
First note that the clique $\set{x^{uv}_1, \dots, x^{uv}_{p-3}, a^{uv}, b^{uv}, c^{uv}}$ needs exactly $\binom{p}{2}$ distinct colors. W.l.o.g. $a^{uv}b^{uv}$ is colored $1$ and the colors used  for this clique are $1$, $2$, \dots, $\binom{p}{2}$. 
None of these colors can be used to color the $r$ edges of the form $d^{uv}y^{uv}_i$ hence they must be colored with $\binom{p}{2} +1 $, \dots, $\binom{p}{2} +r $.
One can observe that an edge $e^{uv}s^{uv}_i$  cannot have a color among  $\binom{p}{2} +1 $, \dots, $\binom{p}{2} +r $ as it is at distance $2$ from the edges of the form $d^{uv}y^{uv}_j$ ($j\in \set{1,\dots,r}$). Moreover this edge cannot receive the same color as one of the edges of the clique $\set{x^{uv}_1, \dots, x^{uv}_{p-3}, a^{uv}, b^{uv}, c^{uv}}$ except  for the color $1$ on the edge $a^{uv}b^{uv}$. Hence all edges of the form $e^{uv}s^{uv}_i$ have the same color.

Now suppose we have a coloring $\gamma$ such that theses edges $e^{uv}s^{uv}_i$ ($i \in \set{1,\dots, 2\ell})$ are all colored with the same color, say~$1$. We color $a^{uv}b^{uv}$ with color~$1$ and use the $\binom{p}{2} + r -1$ other colors to color the rest of the edges of the clique $\set{x^{uv}_1, \dots, x^{uv}_{p-3}, a^{uv}, b^{uv}, c^{uv}}$ and the edges of the form $d^{uv}y^{uv}_j$ ($j\in \set{1,\dots,r}$). We color $e^{uv}z $ for $z \in \set{x^{uv}_1, \dots, x^{uv}_{p-3}, c^{uv}}$ with the color of $a^{uv}z$.

If $r = 0$, then $E_{uv}$ is colored and $\gamma$ is an injective $k$-edge-coloring.

If $r > 0$, we color $d^{uv}e^{uv}$ and $d^{uv}a^{uv}$ with the color of $d^{uv}y^{uv}_1$. We color $d^{uv}z$  for $z \in \set{x^{uv}_1, \dots, x^{uv}_{p-3}, b^{uv}}$ with the color of $c^{uv}z$.  It is left to color the edges of the clique $\set{y^{uv}_1, \dots, y^{uv}_{r}}$, for which we have available the $\binom{p-1}{2}$ colors used to color the clique $\set{x^{uv}_1, \dots, x^{uv}_{p-3}, a^{uv}, b^{uv}}$, which is enough as $r \leq p-1$. This is an injective $k$-edge coloring of $E_{uv}$.
\end{subproof}

Suppose there is an injective $k$-edge-coloring $\gamma$ of $G'$. For an edge $uv$ of $G$, we color it with the color $\gamma(e^{uv}s_1^{uv})$. Take two adjacent edges of $G$: $uv_1$ and $uv_2$. In $S_u$, there is an edge between $v_{1,i}$ and $v_{2,j}$ for some indices $i$ and $j$. Thus the edges $e^{uv_1}v_{1,i}$ and $e^{uv_2}v_{2,j}$ receive different colors. By Claim~\ref{cl:big-k:edge}, $uv_1$ and $uv_2$ receive different colors. Hence $G$ admits a $k$-edge-coloring.

Suppose there is a $k$-edge coloring $\gamma$ of $G$. For each edge $uv$, we color $e^{uv}s_i^{uv}$ with the color $\gamma(uv)$. 
By Claim~\ref{cl:big-k:edge}, we can extend this coloring to all $E_{uv}$. At this point there is no conflict between the colored edges. Indeed the only pairs of edges which are at distance $2$ and not in the same edge gadget are of the form $e^{uw}s_i^{uw}$, and since $\gamma$ is proper, there is no conflict here. It is left to color the edges inside the vertex gadget. Let $e = v_{i,j}v_{i',j'}$ be an uncolored edge. As the maximum degree of the vertices of $S_u$ is at most $\frac{k}{\ell} +2$, there are at most $(\frac{k}{\ell} + 2)^2$ edges incident to a vertex of $S_u$ that can be in conflict with $e$. We must also consider the edges incident with $e^{uv_i}$ and $e^{uv_j}$. For each of the two vertices there is one forbidden color $\gamma(uv_i)$ which is common to $2\ell$ edges incident to $e^{uv_i}$ to which we need to add $p-1$ colors for the other edges of $e^{uv_i}$. In the end, there are at most $2p + (\frac{k}{\ell} + 2)^2$ forbidden colors for $e$. As $2p + (\frac{k}{\ell} + 2)^2 \leq 2p + (\frac{p-1}{4} + 2)^2 = (\frac{p-1}{4})^2 +3p +3 \leq k$ when $k \geq 45$ and $p \geq 10$, $G'$ admits an injective $k$-edge-coloring.
\end{proof}

\section{Conclusion}
We proved that \InjPbName{3} and \InjPbName{4} are NP-complete on some restricted classes of subcubic graphs. 
One can ask whether \InjPbName{5} is NP-complete on subcubic graphs. A conjecture proposed by Ferdjallah \textit{et al.}~\cite{Raspaud19} states that every subcubic graph admits an injective $6$-edge-coloring (it is proved for planar graphs in~\cite{Kostochka20}).
In fact, we only know of two connected subcubic graphs which require six colors: $K_4$ and the prism. Perhaps these are the only examples that are not $5$-colorable, in which case \InjPbName{5} would be polynomial-time solvable for this class.

%\begin{conjecture}
%Every subcubic graph $G$ with no component isomorphic to $K_4$ or the prism admits an injective $5$-edge-coloring.
%\end{conjecture}

% And we conclude with the following, easier, question.

%\begin{question}
%Can it be decided in polynomial-time whether a subcubic graph admits an injective $5$-edge-coloring ?
%\end{question}

We have also proved that for planar bipartite subcubic graphs, \InjPbName{3} is polynomial-time solvable when the girth is at least~$16$ (because the answer is always YES), but NP-Complete when the girth is~$6$. It would be interesting to determine the values of the girth of planar bipartite subcubic graphs for which \InjPbName{3} stays NP-Complete, becomes polynomial-time solvable, and always has YES as an answer.

We also do not know whether \InjPbName{4} is NP-Complete for bipartite subcubic graphs.

\section{Acknowledgments}
This research was supported by the IFCAM project ``Applications of graph homomorphisms'' (MA/IFCAM/18/39) and by the ANR project HOSIGRA (ANR-17-CE40-0022).

\bibliographystyle{plain}
\bibliography{biblio}
\end{document}